\documentclass[twocolumn,superscriptaddress,preprintnumbers,amsmath,amssymb,prx]{revtex4-1}

\newcommand{\IMSS}{Muon Science Laboratory and Condensed Matter Research Center, Institute of Materials Structure Science, High Energy Accelerator Research Organization (KEK-IMSS), Tsukuba, Ibaraki 305-0801, Japan}
\newcommand{\Sokendai}{Department of Materials Structure Science, The Graduate University for Advanced Studies (Sokendai), Tsukuba, Ibaraki 305-0801, Japan}
\newcommand{\TRIUMF}{Centre for Molecular and Materials Science TRIUMF, Vancouver, B.C. V6T 2A3, Canada}
\newcommand{\SagaU}{Graduate School of Science and Engineering, Saga University, Saga 840-8502, Japan}
\newcommand{\NIMS}{National Institute for Materials Science (NIMS), Tsukuba, Ibaraki 305-0044, Japan}
\newcommand{\ChuoU}{Department of Physics, Chuo University, Tokyo 112-8551, Japan}

\usepackage{txfonts}
\usepackage[dvipdfmx]{graphicx}
\usepackage{dcolumn} 
\usepackage{bm} 
\usepackage{xspace}
\usepackage{ulem} 
\usepackage{multirow}
\usepackage{tabularx}
\usepackage{breqn}
\usepackage[hypertex,colorlinks=true,linkcolor=black,citecolor=blue,filecolor=blue,urlcolor=blue,setpagesize=false,nesting=true]{hyperref}
\begin{document}

\author{H.~Okabe}
\affiliation{\IMSS}
\author{M.~Hiraishi}
\affiliation{\IMSS}
\author{A.~Koda}
\affiliation{\IMSS}\affiliation{\Sokendai}
\author{S.~Takeshita}
\affiliation{\IMSS}\affiliation{\Sokendai}
\author{K.~M.~Kojima}
\affiliation{\TRIUMF}
\author{I.~Yamauchi}
\affiliation{\SagaU}
\author{T.~Ohsawa}
\affiliation{\NIMS}
\author{N.~Ohashi}
\affiliation{\NIMS}
\author{H.~Sato}
\affiliation{\ChuoU}
\author{R.~Kadono}
\thanks{Corresponding author: ryosuke.kadono@kek.jp}
\affiliation{\IMSS}\affiliation{\Sokendai}


\title{Local electronic structure of dilute hydrogen in  $\beta$-MnO$_{2}$}
\begin{abstract}
The electronic and magnetic states of $\beta $-MnO$_{2}$ in terms of hydrogen impurities have been investigated by muon spin rotation ($\mu$SR) technique combined with density-functional theory (DFT) calculations for muon as pseudo-hydrogen. We found that 85 {\%} of implanted muons are localized in the oxygen channels of the rutile structure and behave as interstitial protons (Mu$^{+})$ except those (7.6{\%}) forming a charge-neutral state (Mu$^{0})$ at 2.3 K, which indicates that interstitial hydrogen acts as a shallow donor within less than 0.1 meV of ionization energy. The residual 15{\%} of muons are attributed to those related to lattice imperfection as Mn vacancies. Detailed analyses combined with DFT approach suggested that the muon is localized at the center of the oxygen channel due to its large zero-point vibration energy. 
\end{abstract}

\maketitle

\section{Introduction}
Manganese dioxide (MnO$_{2})$ is well known as a catalyst for a variety of chemical reactions such as the deoxidization process of hydrogen peroxide, with which we are familiar in elementary school chemistry, or most recently, it is attracting a lot of attention as a highly efficient heterogeneous catalyst for the oxidation of various types of substrates including bio-mass-derived compounds ~\cite{cat}. The most known applications of MnO$_{2}$ are in batteries; MnO$_{2}$ synthesized by the electrolytic method ($\gamma $-MnO$_{2})$ is used worldwide as the cathode material in lithium or zinc primary battery ~\cite{bat}. The artificially synthesized MnO$_{2}$ generally contains hydrogen in the crystal structure, which are supposed to influence various physical, chemical, and electrochemical properties.
\par
There are two types of hydrogens (protons) existing in $\gamma $-MnO$_{2}$ proposed by Ruetschi and Giovanoli ~\cite{Rucy1,Rucy2,Colm}. One is that protons associated with cation vacancies, termed "Ruetschi'' protons, which compensate for the charge instead of Mn$^{4+}$ cations. The other is that protons associated with Mn$^{3+}$ cations, termed "Coleman'' protons, which are located in the tunnels of the structure and more mobile than Ruetschi protons. The existence of the two types of protons has been confirmed experimentally by neutron diffraction and NMR studies ~\cite{Filax,JTES,Pitel,JMS,NMR}. However, these protons are mainly distinguished by differences in O-H bond lengths, therefore its specific positions and electronic states remain unclear. The main reason for the problem is the difficulty of detecting dilute amounts of hydrogen.
\par
Since hydrogen is the lightest and smallest of all the elements, and ubiquitous presence in nature, hydrogen easily penetrates into materials to cause embrittlement of metals, or serves as possible sources for unintentional carrier doping in semiconductors ~\cite{Van1,Van2}. Nevertheless, it is difficult to find out the behavior of dilute hydrogen in matter due to its own weak signals and/or false detections of background-originated signals from extraneous hydrogen in various diffraction analyses. It is therefore desirable to develop an effective method to ascertain the behavior of hydrogen in host materials.
\par
To obtain microscopic information on the local electronic structure of isolated hydrogen in matter, positive muon studies have been making significant contributions. While positive muon as pseudo-hydrogen exhibits a remarkable isotope effect for kinetics of muon/proton in matter, its local electronic structure as an atom determined by muon-electron interaction (resulting in various charged states designated as Mu$^{+}$, Mu$^{0}$, or Mu$^{-})$ is almost identical with that of hydrogen. Thus, implanted muon can be regarded as a simulator for the electronic structure of interstitial hydrogen. We previously identified the electronic structure of muons in oxide semiconductors using muon spin rotation ($\mu $SR) technique, and found evidence for hydrogen impurities as a possible origin of unintentional $n$-type conductivity ~\cite{Gil,Cox,Shimo1,Salman,Shimo2,Vilao,C12A7,FeS2,IGZO}.
\par
As a part of our continued effort to elucidate the behavior of hydrogen in oxide semiconductors, we focused on $\beta $-MnO$_{2}$ as our next research target. MnO$_{2}$ is known to exhibit a variety of crystalline polymorphs, which results in various tunnel and layered structures ~\cite{polym}, where $\beta $-MnO$_{2}$ (pyrolusite) is the simplest and thermodynamically stable phase. The crystal structure of $\beta $-MnO$_{2}$ consists of MnO$_{6}$ octahedra shared by corners or edges with 1 $\times$ 1 tunnels (oxygen channels, see Fig. 1), which is also part of the basic structure of $\gamma $-MnO$_{2}$. It would be interesting to note that the structure of $\beta $-MnO$_{2}$ is identical with rutile (TiO$_{2})$, which we have intensively studied by $\mu $SR ~\cite{Shimo2,Vilao}.
\par
$\beta -$MnO$_{2}$ is regarded as a narrow gap semiconductor and it is known to exhibt a $n$-type conductivity ~\cite{Noda,Sato}. The $n$-type nature of $\beta $-MnO$_{2}$ may stem from oxygen deficiency and/or lesser impurity interstitials like hydrogen. If oxygen vacancy concentration is negligibly small, hydrogen could be a major source of carriers. However it is often difficult to separate all these different contributions in the bulk property measurements. On the other hand, $\mu $SR can provide useful information to extract intrinsic electronic properties out of those extrinsic contributions because muon monitors local magnetic states in the atomic scale. Even more conveniently, $\beta $-MnO$_{2}$ is known as a classical example of magnetic materials with a well-defined screw magnetic order below $T_{N}  =$ 92 K ~\cite{SatoX,Regls}. Since muon is a direct probe of local magnetic fields, we can determine the position and electronic state of hydrogen in $\beta $-MnO$_{2}$ by $\mu $SR, using the magnetic structure as a guide.
\par
In this work, we used muon as a microscopic simulator for extrinsic hydrogen/proton to identify the local electronic and magnetic structures stemmed from interstitial hydrogen in $\beta $-MnO$_{2}$, by probing the local fields from surrounding Mn ions that exhibit the well-defined magnetic order. The paper proceeds as follows. Section II describes the sample characterization, $\mu $SR experimental, and computational details. Section III presents the results of transverse-field (TF) $\mu $SR, zero-field (ZF) $\mu $SR measurements, and density functional theory (DFT) calculations. Then, we discuss the positions of hydrogen and the effect on the electronic and magnetic state of $\beta $-MnO$_{2}$. Finally, Sec. IV summarizes this paper.

\begin{figure}[!t]
	\centering
	\includegraphics[width=0.85\linewidth,clip]{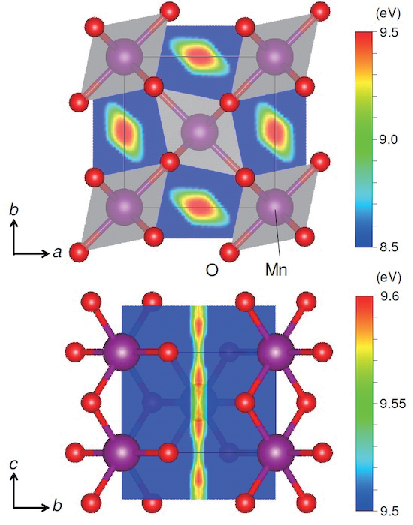}
	\caption{Crystal structure and electrostatic potential maps for $\beta $-MnO$_{2}$ calculated by using reported lattice parameters ~\cite{Baur}. The Hartree potential calculated using VASP are represented by the color contour from blue to red in the lattice planes (002) and (100). The solid lines indicate the unit cell.}
	\label{Fig1}
\end{figure}
\section{Experimental Details}
Powder sample of $\beta $-MnO$_{2}$ (3N, Soekawa Co., Ltd.) was used in the present study. The crystal structure and phase purity were checked using a Rigaku MultiFlex powder X-ray diffractometer under ambient conditions with Cu K$\alpha $ radiation. The XRD data indicate that the sample crystallized into $\beta $-MnO$_{2}$ single phase and the absence of secondary phase (see Supplemental Material Fig. S1).
\par
The hydrogen content in the sample was observed by thermal desorption spectrometry (TDS) placed in the National Institute for Material Science (NIMS). The sample was linearly heated from room temperature to 1073 K in 2 hours and kept at 1073 K for 1 hour. During the heating, the amount of desorbed hydrogen was measured by a quadrupole mass spectrometer (QMS). The QMS signal intensity of hydrogen molecule from sample increase with increasing temperature, and 2.918 $\times$ 10$^{19\, }$cm$^{-3}$ of hydrogen was detected (Fig. S2).
\par
The magnetic susceptibility $\chi $ was measured using a Quantum Design SQUID magnetometer from 5 to 400 K. The temperature dependence of $\chi $ shows a kink at 92 K, corresponding to the N\'{e}el temperature of $\beta $-MnO$_{2}$ (Fig. S3). The peak anomaly around 40 K is probably due to hydrogen impurities, since it was suppressed after heat treatment at 723 K in air. $\chi $ also shows the Curie-Weiss behavior with the Curie constant $C =$ 2.51 emu k/mol and the Weiss temperature $\Theta  = -$788 K, which are in good agreement with the earlier report ~\cite{Sato} (Fig. S3, inset).
\par
Conventional $\mu $SR experiments were performed using the ARTEMIS spectrometer installed in the S1 area at Muon Science Establishment (MUSE), Japan Proton Accelerator Research Complex (J-PARC). Time dependent $\mu $-e decay asymmetry A($t)$ was measured to monitor the local field under zero field (ZF), longitudinal field (LF), and weak transverse field (TF) conditions. Additional $\mu $SR experiments were conducted to resolve the detailed local field distribution using two instruments, i.e., the LAMPF spectrometer on the M20 beamline and NuTime spectrometer on the M15 beamline at TRIUMF, Canada, for the magnetic and paramagnetic phases of $\beta $-MnO$_{2}$, respectively. The $\mu $SR spectra have been analyzed using the MUSRFIT software package ~\cite{Suter}.
\par
The DFT calculations were performed using the projector augmented wave approach ~\cite{PAW} implemented in the Vienna ab initio simulation package (VASP) ~\cite{VASP} with the Perdew-Burke-Ernzerhof (PBE) exchange correlation potential ~\cite{PBE}. The cutoff energy for the plane-wave basis set was 500 eV. All atoms were relaxed until the Hellmann$-$Feynman forces on them were smaller than 0.01 eV/{\AA}. The distribution of the local magnetic field at the muon sites was calculated using Dipelec program ~\cite{Koji}. Crystal structures were visualized using the VESTA program ~\cite{Izumi}.
\section{Results and discussion}
\subsection{Local fields in the paramagnetic phase}
At first, we attempted to narrow down the position of hydrogen/muon in $\beta $-MnO$_{2}$ structure by examining the local field in the paramagnetic phase. The local field provides useful information for identifying the muon stopping site, because it is mainly determined by the spatial arrangement of the nearest neighboring (nn) Mn $d$-electrons.
As shown in Fig. 1, it is suggested from our preliminary calculation using VASP that the Hartree potential for the interstitial Mu$^{+}$ exhibits minima around the center of the oxygen channels along the $c$-axis direction. When an external magnetic field is applied transverse to the initial muon polarization vector, the muon exhibits spin precession at a frequency corresponding to the summation of the external and internal fields. Thus, by evaluating the shift from the reference frequency of the external field, one can extract the contribution of the internal field at a particular stopping site from the TF-$\mu $SR spectrum.
\par
Figure 2 shows the fast Fourier transform (FFT) of TF-$\mu $SR spectra under a transverse field of 6 T obtained at various temperatures. One can see a single Gaussian-like lineshape at around 813 MHz, where the slight broadening of linewidth with decreasing temperature is probably due to the gradual development of staggered magnetization related to the short range magnetic order ~\cite{Sato,Regls}. To interpret the lineshape of these spectra, we calculated frequency distribution in polycrystalline sample (powder pattern) for several candidate muon sites ~\cite{Slichter}.
\par
\begin{figure}[!t]
	\centering
	\includegraphics[width=\linewidth,clip]{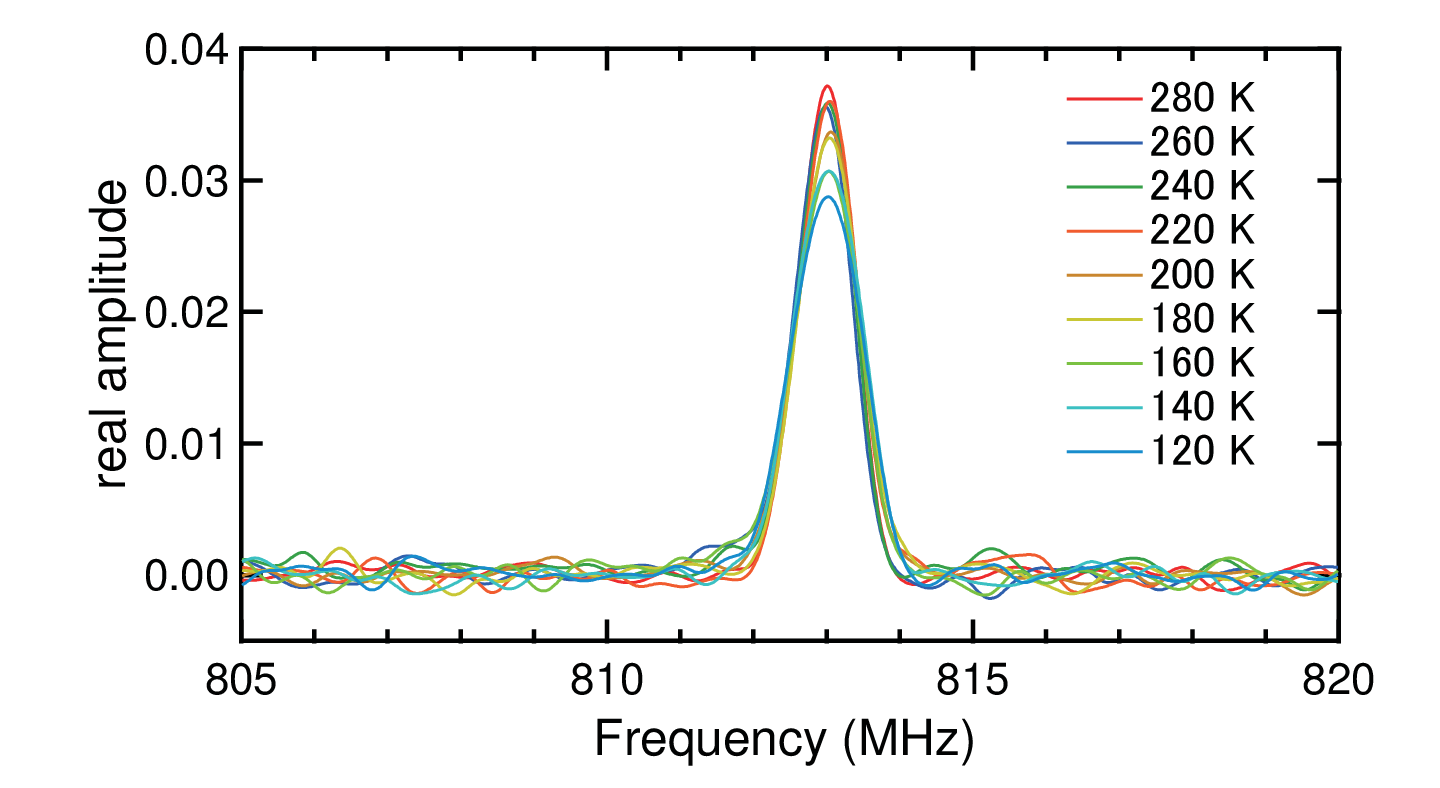}
	\caption{Fast Fourier transform of the $\mu $SR time spectra measured under a transverse field of 6 T.}
	\label{Fig2}
\end{figure}
\par
The calculated powder patterns of putative muon sites are shown in Fig. 3 with additional Gaussian broadening due to the local field inhomogeneity and  the truncation of Fourier transforms by the finite time window ($ t \leq$ 10 $\mu$s). These muon sites have been selected by considering the electrostatic potential maps (see Fig. 1) and the distance from nn O ions. The lineshape has two edges determined by the anisotropy of the hyperfine coupling (A$_{\parallel}$ and A$_{\mathrm{\bot }})$, which is related with the local $\chi $ induced by the nn Mn spins. One can see that the muon site near a certain Mn ion, $e.g.$ 2$b$ site (0, 0, 1/2), shows a marked asymmetric feature with broad distribution. Since the experimental spectra show no signs of these shoulder-like features, implanted muons tend to be located away from Mn ions, at a high symmetry position in the Mn sublattice. Comparison of the spectra with that simulated for the O defect position (vacancy) also infers that muons have little chance to find O vacancies within their lifetime.
\par
\begin{figure}[!t]
	\centering
	\includegraphics[width=\linewidth,clip]{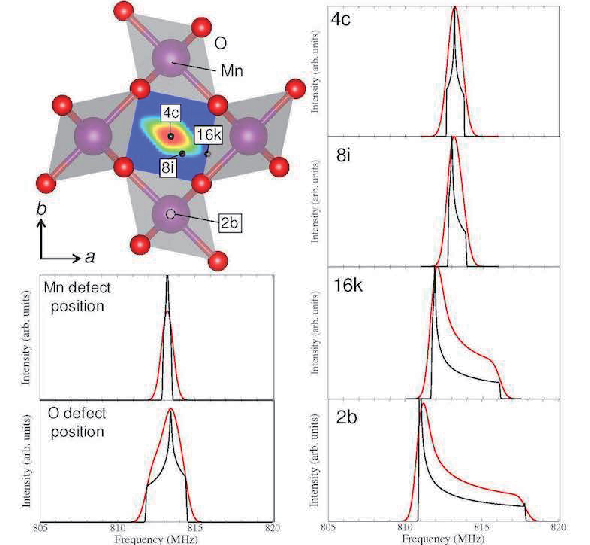}
	\caption{Simulated powder patterns for the putative muon stopping sites in $\beta $-MnO$_{2}$ under a transverse field of 6 T. Each interstitial crystallographic position is represented in a Wycoff letter: 4$c$ (1/2, 0, 0), 8$i$ (0.42667, 0.11185, 0), 16$k$ (0.48954,0.26299, 0.18754), 2$b$ (0, 0, 1/2). The black lines represent the calculated powder patterns. The red lines represent the additional Gaussian broadening (see the main text) superimposed on the powder patterns.}
	\label{Fig3}
\end{figure}
\par
Considering previous neutron and NMR studies ~\cite{Filax,JTES,Pitel,JMS,NMR}, it is reasonable to assume that muons are localized near the center of the oxygen channel in the rutile structure. Our result further suggests that muons do not occupy low symmetry positions including oxygen vacancies. On the other hand, there remains a possibility that muons are localized in Mn vacancies due to the lack of frequency resolution (see the lineshape at the Mn defect position in Fig. 3). Therefore, it requires additional information to resolve this issue (see the next section).
\subsection{Local fields in the magnetically ordered phase}
\par
\begin{figure}[!t]
	\centering
	\includegraphics[width=\linewidth,clip]{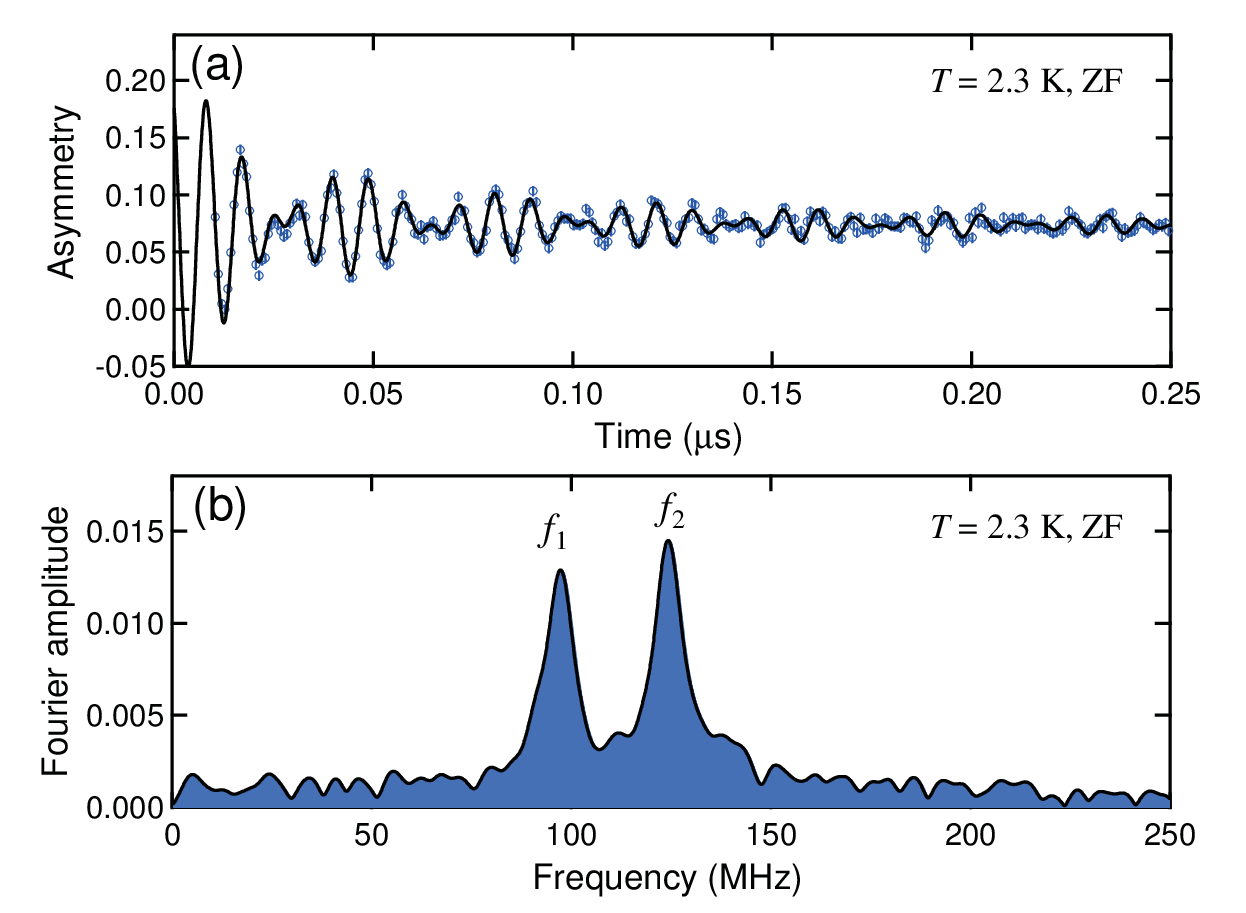}
	\caption{(a) $\mu $SR time spectrum at 2.5 K under a zero external field. The solid line represents the result of least-squares fitting (see the main text). (b) Fast Fourier transform of the time spectrum.$ f_{1}$ ($f_{2})$ indicates a peak frequency.}
	\label{Fig4}
\end{figure}
\par
\begin{figure}[!t]
	\centering
	\includegraphics[width=\linewidth,clip]{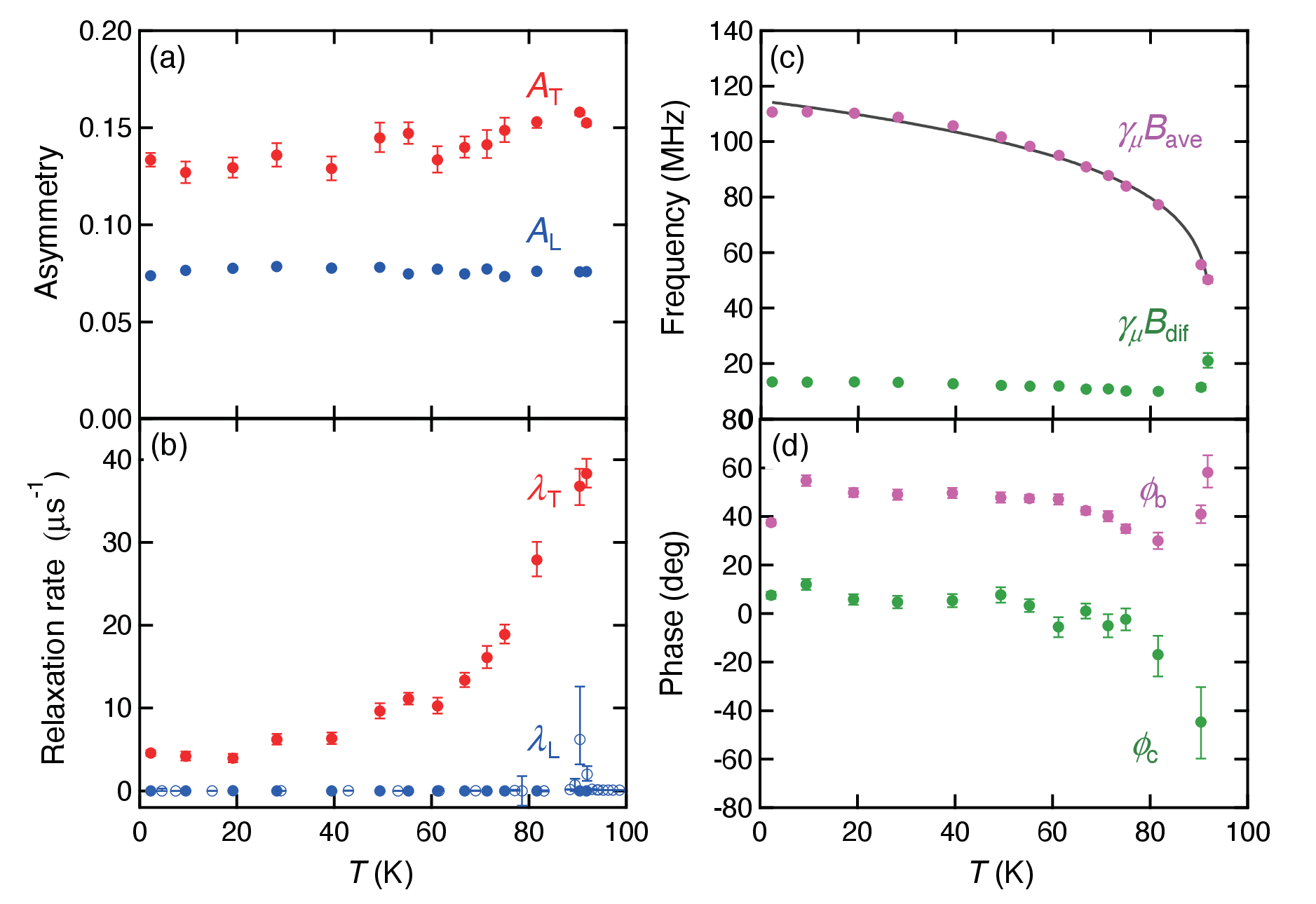}
	\caption{Temperature dependence of (a) the transverse (longitudinal) asymmetry $A_{T}$ ($A_{L})$, (b) the transverse (longitudinal) relaxation rate $\lambda _{T}$ ($\lambda_{L})$, (c) the precession frequencies $\gamma_{\mu }B_{\mathrm{ave}}$ and $\gamma_{\mu }B_{\mathrm{dif}}$, and (d) the initial phase $\varphi_{b}$ and $\varphi_{c}$. The open circle in (b) indicates $\lambda_{L}$ obtained at J-PARC MUSE. The solid line in (c) indicates the result of least-squares curve-fit by the power-law function (see the main text).}
	\label{Fig5}
\end{figure}
\par
$\beta$-MnO$_{2}$ undergoes a helical magnetic transition at $T_{N}  =$ 92 K with incommensurate magnetic modulation [propagation vector $q_{m}  =$ (0, 0, $\sim$2/7)] without much temperature dependence ~\cite{SatoX, Regls}. Given that the magnetic structure is already known, a spontaneous long-range magnetic order is often helpful to identify the muon sites by comparing the observed field distribution with that calculated for the candidate sites using the known magnetic structure.
\par
Figure 4(a) shows a ZF-$\mu $SR time spectrum at 2.5 K for the sample from the same batch used in the TF measurements. One can see a clear oscillation in the early time domain (\textless 0.25 $\mu $s), implying the long-range magnetic order. The FFT spectrum [Fig. 4 (b)] indicates that the oscillating signal is dominated by two frequency components ($f_{1\, }\simeq $ 97 and $f_{2\, }\simeq $ 124 MHz). While this feature is usually attributed to the presence of two distinct muon sites, it turns out to be not necessarily the case in $\beta $-MnO$_{2}$.
\par
For an incommensurate magnetic structure under zero field, the local field distribution $D(B_{\mathrm{loc}})$ is given by
\par
\[
D\left( B_{\mathrm{loc}} \right)=\frac{2}{\pi }\frac{B_{\mathrm{loc}}}{\sqrt 
{B_{\mathrm{loc}}^{2}-B_{\mathrm{min}}^{2}} \sqrt 
{B_{\mathrm{max}}^{2}-B_{\mathrm{loc}}^{2}} }\, ,
\]
\par
\noindent
where it has two characteristic cutoff field ($B_{\mathrm{min}}$ and $B_{\mathrm{max}})$~\cite{Yao}. Unfortunately, the muon polarization function in time domain associated with $D(B_{\mathrm{loc}})$ cannot be obtained analytically, we adopt an approximate distribution which is called shifted-Overhauser distribution$ D'(B_{\mathrm{loc}})$,
\par
\[D'\left( B_{\mathrm{loc}} \right)=\frac{1}{\pi }\frac{1}{\sqrt {B_{\mathrm{dif}}^{2}-\left( B_{\mathrm{loc}}-B_{\mathrm{ave}} \right)^{2}} }\, ,\]
\par
\noindent
where $B_{\mathrm{dif}}=$ ($B_{\mathrm{max}}- B_{\mathrm{min}})$/2 and $B_{\mathrm{ave}}=$ ($B_{\mathrm{max}} + B_{\mathrm{min}})$/2, is used for the incommensurate case ~\cite{Amato}. According to this approximation, the oscillatory part of the muon polarization function is written by the product of the zeroth order Bessel function and the cosine function.
\par
Based on the above approximation, we analyzed the spectrum by following curve-fit function,
\par
\begin{dmath*}
A\left( t \right)=A_{T}J_{0}\left( \gamma_{\mu }B_{\mathrm{dif}}t+\phi _{\mathrm{b}} \right)\cos {\left( \gamma_{\mu }B_{\mathrm{ave}}t+\phi_{c} \right)\mathrm{exp}\left( -\lambda_{T}t \right)}
+A_{L}\mathrm{exp}\left( -\lambda_{L}t \right)
\end{dmath*}
\par
\noindent
where the amplitude $A_{T}$ ($A_{L})$ is proportional to the fraction perpendicular (parallel) to initial polarization direction, $J_{0}$ is a Bessel function of the first kind, $\gamma_{\mu \, }(=$ 2$\pi $ $\times$ 135.53 MHz/T) is the muon gyromagnetic ratio, $\varphi_{b}$ ($\varphi_{c})$ is the initial phase of each function, respectively, and $\lambda_{T}$ ($\lambda_{L})$ is the transverse (longitudinal) relaxation rate. The solid curve in Fig. 4(a) shows the result of least-squares curve-fit. The temperature dependence of respective parameters below $T_{N}$ are shown in Figs. 5.
\par
The slight decrease of $A_{T}$ at lower temperatures is attributed to the partial loss of initial muon polarization (which is discussed below). The development of the magnetic order parameter with decreasing temperature manifests in the evolution of the frequency $\gamma_{\mu }B_{\mathrm{ave}}$ as shown in Fig. 5(c). The temperature dependence of $\gamma_{\mu }B_{\mathrm{ave}}$ shows a critical behavior described by the power law, $\gamma _{\mu }B_{\mathrm{ave}}(T)\mathrm{\, \propto \, }(T_{N}-T)^{\beta }$ with $T_{N}  =$ 92.7(2) K and $\beta =$ 0.182(9). The value of critical exponent $\beta $ is in good agreement with those obtained from the synchrotron x-ray magnetic scattering ~\cite{SatoX} and the neutron scattering ~\cite{Regls}. Although the phase parameter $\varphi_{c}$ is nearly zero except around $T_{N}$, $\varphi_{b}$ tends to be shift by $\sim$ $\pi $/4 compared with $\varphi_{c}$. It may be related to the $\pi $/4 phase shift between the Bessel function and the cosine function ~\cite{Yao}.
\par
The temperature dependence of the observed signal fraction, [{$A_{T}+A_{L}$]/$A_{0}$ (red circle) and the longitudinal fraction, $A_{L}$/$A_{0}$ (blue circle) are shown in Fig. 6, where $A_{0}$ ($=$ 0.23) is estimated from the full asymmetry above $T_{N}$. One can clearly see that the signal fraction below $T_{N}$ is less than 90 {\%} of the total volume fraction. In contrast, the longitudinal fraction ($z$ component) is almost one third of the total volume fraction, indicating that all the implanted muons are subject to isotropic static magnetic order. These facts indicate that there is an unresolved signal component (missing fraction) which is depolarized within the dead time ($t$ \textless 0.01 $\mu $s) of the experiment. 
\par
The missing fraction at 2.3 K is estimated to be about 15 {\%} of the total volume fraction. Such a fast depolarization is attributed to the strong disorder of $D(B_{\mathrm{loc}})$. Considering that the possibility of muons occupying O vacancy is unlikely as inferred from $D(B_{\mathrm{loc}})$ (at least above $T_{N}$, see Fig.3), it is reasonable to attribute the origin of the missing fraction to the muons occuping Mn vacancies that corresponds to naturally included Ruetschi protons. This also leads to the hypothesis that the residual 85 {\%} of muons corresponds to interstitial hydrogens (Coleman protons) which occupies the oxygen channels. 
\par
In close examination of the temperature dependence of $A_{L}$/$A_{0}$ in Fig. 6, one can find that it slightly declines from the one third line as temperature decreases. This is attributed to the formation of paramagnetic Mu$^{0}$ state which is subject to spin/charge exchange process with diamagnetic state (Mu$^{+}  +$ e$^{-}$ $\to$ Mu$^{0})$. The reduction of $A_{L}$ strongly suggests that the electronic state bound to the interstitial muon is not polaronic (i.e., the electron spin not being locked along the Mn spins), but rather atomic to allow spin-triplet ($F=1$) or spin-singlet ($F= 0$) state relative to the muon spin direction. The precession frequency for the $F = 0$ state usually exceeds the limit determined by experimental time resolution ~\cite{Schenck}. Therefore, only residual polarization for the $F = 1$ state can be observed, resulting in a reduction of the longitudinal fraction by half. If Mu$^{0}$ is present, the longitudinal asymmetry $A_{L}$ is reproduced by following linear combination
\par
\[A_{\mathrm{L}}=\frac{1}{3}\left( 1-{\frac{1}{2}f}_{\mathrm{Mu}} \right)A_{\mathrm{0}},\]
\par
\noindent
where $f_{\mathrm{Mu}}$ is the fractional yield of Mu$^{0}$.
\par
The temperature dependence of longitudinal signal fraction $f_{\mathrm{L}}$ (see Fig.6 inset) is adequately described by the thermal activation model $f_{\mathrm{L}}$ = $f_{\mathrm{d}}$exp$(-E_{\mathrm{a}}$/2$k_{\mathrm{B}}$$T$), where $f_{\mathrm{d}}$ is the diamagnetic longitudinal fraction ($f_{\mathrm{L}}$ = $f_{\mathrm{d}}$, when $f_{\mathrm{Mu}}$ = 0),  $E_{\mathrm{a}}$ is the characteristic energy, and $k_{\mathrm{B}}$ is Boltzmann's constant. A fit of $f_{\mathrm{L}}$ (blue line in the inset of Fig.6) below 50 K yields $f_{\mathrm{d}}$ = 0.3420(8) and $E_{\mathrm{a}}$ = 0.0292(8) meV/K. Considering the value of $f_{\mathrm{L}}$ = 0.316(1) at 2.3 K, $f_{\mathrm{Mu}}$ is estimated to be 0.0760(4) (which may be a lower bound for the Mu$^{0}$ yield). This means that the electrons from the hydrogen impurity levels are mostly promoted to the conduction band in most temperature ranges. In other words, the donor level associated with hydrogen is estimated to be extremely shallow, probably less than 0.1 meV. Since the diamagnetic muons in the oxygen channels (corresponding to the Coleman proton) can be attributed to Mu$^{0}$ in its ionized state, they are presumed to serve as electron donors. For these reasons, the interstitial hydrogen is expected to be a source of $n$-type conductivity in $\beta $-MnO$_{2}$. We note that the hydrogen concentration (2.918 $\times$ 10$^{19\, }$cm$^{-3})$ obtained from our TDS measurement is comparable with the carrier density ($n =$ 4.8 $\times$ 10$^{19\, }$cm$^{-3})$ reported in the previous work ~\cite{Sato}, which is perfectly in line with this scenario.
\par
\begin{figure}[!t]
	\centering
	\includegraphics[width=\linewidth,clip]{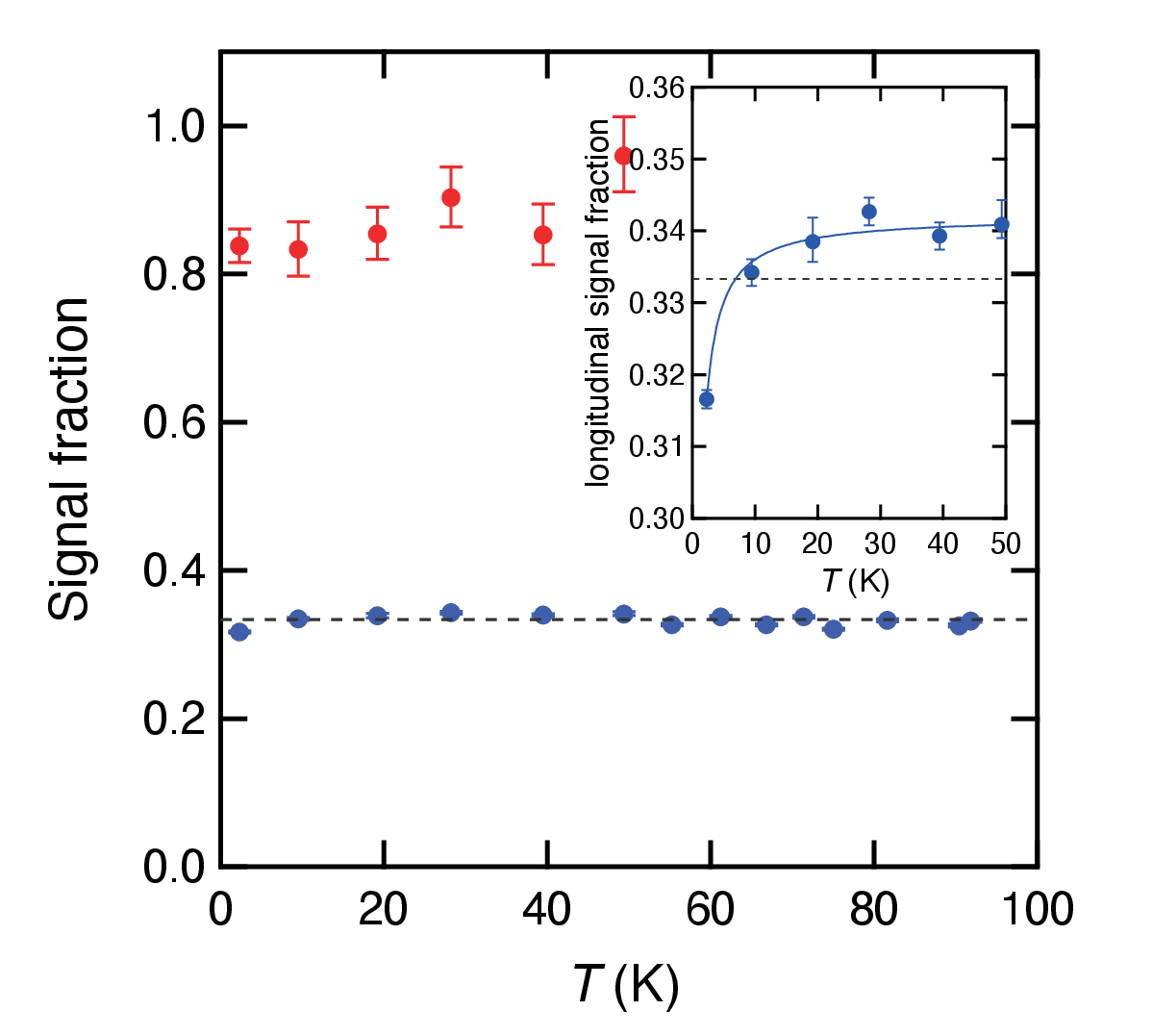}
	\caption{Temperature dependence of the observed signal fraction (red circle) and the longitudinal signal fraction $f_{\mathrm{L}}$ (blue circle). The inset shows the enlargement below 50 K with the fitting curve using the thermal activation model (see main text). The broken line represents the 1/3 of the total fraction.}
	\label{Fig6}
\end{figure}
\par
\subsection{First-principles calculations}
In order to gain more understanding of hydrogen states in $\beta $-MnO$_{2}$, the DFT calculations were performed using VASP for comparisons with the $\mu $SR experiments. As the first step, we calculated the local field  $B_{\mathrm{loc}}$ at the most energetically stable position 4$c$ (1/2, 0, 0) with the screw periodicity $q =$ 2$c$*/7 and Mn$^{4+}$ moment size of 2.34$\mu_{B}$ ~\cite{Regls}. Under zero-external field,  $B_{\mathrm{loc}}$ is determined by a vector sum of the magnetic dipolar field of the Mn ions and the Fermi contact interaction with unpaired electron in the muon 1$s$ orbital ~\cite{Hayano},
\par
\[B_{\mathrm{loc}}=\sum\limits_i {\hat{A}_{i}\mu_{i}} +\frac{8}{3}\pi S_{\mathrm{e}}\delta \left( r \right)\]
\par
\noindent
where $\mu_{i}$ is the magnetic moment of the $i$-th Mn ion located at distance $r_{i} =$ ($x_{i}$, $y_{i}$, $z_{i})$ from the muon site, $S_{\mathrm{e}}$ is the electron spin magnetic moment, and $\delta (r)$ is the delta function, and
\par
\[\hat{{\bm A}_i}=A_i^{\alpha\beta}=\frac{1}{r_i^3}\left(\frac{3\alpha_i\beta_i}{r^2_i}-\delta_{\alpha\beta}\right),\: (\alpha, \beta = x,y,z)\]
\par
\noindent
is the dipolar tensor. The Fermi contact term is basically negligible except for the Mu$^{0}$ state.
\par
Figure 7(b) shows a simulated distribution of  $B_{\mathrm{loc}}$ for the 4$c$ site, which is in remarkable agreement with the experimental result [Fig. 7(a)]. A comparative study with DFT calculation provides useful information for identifying the muon's location and state. For example, if all the nn Mn$^{4+}$ ($S =$ 3/2) ions bear the full moment (3$\mu_{B})$, the calculated distribution of$ B_{\mathrm{loc}}$ becomes broader, leading to the shift of central frequency by $\sim$ 30 MHz (not shown), which is in disagreement with the experimental result. This also supports the moment reduction of Mn ions (2.34$\mu_{B})$ from the viewpoint of local magnetism probed by $\mu$SR.
\par
We further extended the simulation of  $B_{\mathrm{loc}}$ to investigate the influence of structural relaxation by a muon/hydrogen insertion. The Brillouin zone for a 2 $\times$ 2 $\times$ 3 supercell was sampled using a cutoff energy of 500 eV and a 4 $\times$ 4 $\times$ 4 $k$-point mesh for calculations. The structure optimization was performed using lattice parameters and all of the atomic positions were allowed to relax until the forces acting on each ion became less than 0.01 eV/{\AA}. Note that the spin degree of freedom and the electron correlation parameter $U$ are not incorporated, because the difference between optimized atomic positions of Mn around the hydrogen is small irrespective of the presence or absence of $U$(= 4 eV). Therefore, it is not an issue in this study.
\par
The result of calculation is shown in Fig. 8. The hydrogen occupies slightly off-center of the oxygen channel where it forms O-H bond with a bond length of 0.99 {\AA} and donating electron to the host lattice. The surrounding oxygen show large displacement from their initial crystallographic positions due to hydrogen bond. On the other hand, the surrounding cationic sub-lattice of Mn ions shows a slight expansion due to the electrostatic repulsion between H$^{+}$ and Mn$^{4+}$ ion; resulting in 1.4$\sim$7.2{\%} Mn-Mn bonds elongation compared to those without H insertion. Taking into account the displacement of neighboring six Mn ions from H, we have simulated  $B_{\mathrm{loc}}$ using a 686-octahedron 7 $\times$ 7 $\times$ 7 supercell to reproduce the partially distorted $\beta $-MnO$_{2}$ lattice considering the period of the helical magnetic structure.
\par
When muon (or hydrogen) is bound to an oxygen, the donated electron can localize on a neighboring transition metal $d$ orbital to form a charge-neutral complex (i.e., polaron) ~\cite{Shimo2,Vilao,Cr2O3}. In light of this possibility, we simulated  $B_{\mathrm{loc}}$ with two different charge distributions: a polaronic electron distributed on the nn Mn ion only [Fig. 7(c)] or the nearby three Mn ions [Fig. 7(d)], that are not obtained from the DFT calculation. Although subtle differences in shape are observed in comparison with the non-distorted case [Fig. 7(b)], they still have a characteristic two peak structure. Note that when the structure was optimized by fixing hydrogen at the 4$c$ site, the Mn-H distance was uniformly extended for the all Mn ions. Therefore, no significant change in the shape of the local magnetic field was obtained.
\par
Hydrogen generally takes an asymmetric O-H\textperiodcentered \textperiodcentered \textperiodcentered O configuration with a short O-H covalent bond and a long H\textperiodcentered \textperiodcentered \textperiodcentered O hydrogen bond. However, it resides equidistantly between two nearest oxygen atoms under high pressures. This phenomenon is called \textit{H-bond symmetrization} and often observed in high-pressure phase of ice ~\cite{ice}. The H-bond symmetrization is closely linked to zero-point vibration, thus the isotope effect is expected to be more pronounced ~\cite{Hsym,Hira}. For example, $\delta $-AlOOH, which has a distorted-rutile-type framework, shows the H-bond symmetrization under high pressures with a significant isotope effect; the deuteration induces a shift of the symmetrization to a higher pressure ~\cite{AlOOH}. It is commonly known that a lighter isotope possesses a larger zero-point energy. The presumption that muon locates in the middle of two nearest oxygen atoms to form a symmetric O-$\mu $-O bond due to the isotope effect would be a possible scenario to explain the muon localization at the 4$c$ site.
We have evaluated the zero-point motion energy of muon and hydrogen from the potential energy distribution along O-4$c$-O direction by the DFT calculation using VASP (see Supplemental Material Fig. S4). Given the energy difference between the ground state ($E_0$) and the first excited state ($E_1$) of muon or hydrogen, muon can tunnel between the two wells at a very fast timescale as far as the muon spin precession frequency is much lower than the tunneling frequency. This picture can be approximated as a kind of resonant state where muon locates at the center of the tunnel. In such a case where the quantum nature is pronounced, muon may not be able to simulate the behavior of dilute hydrogen defects correctly.
\par
\begin{figure}[!t]
	\centering
	\includegraphics[width=\linewidth,clip]{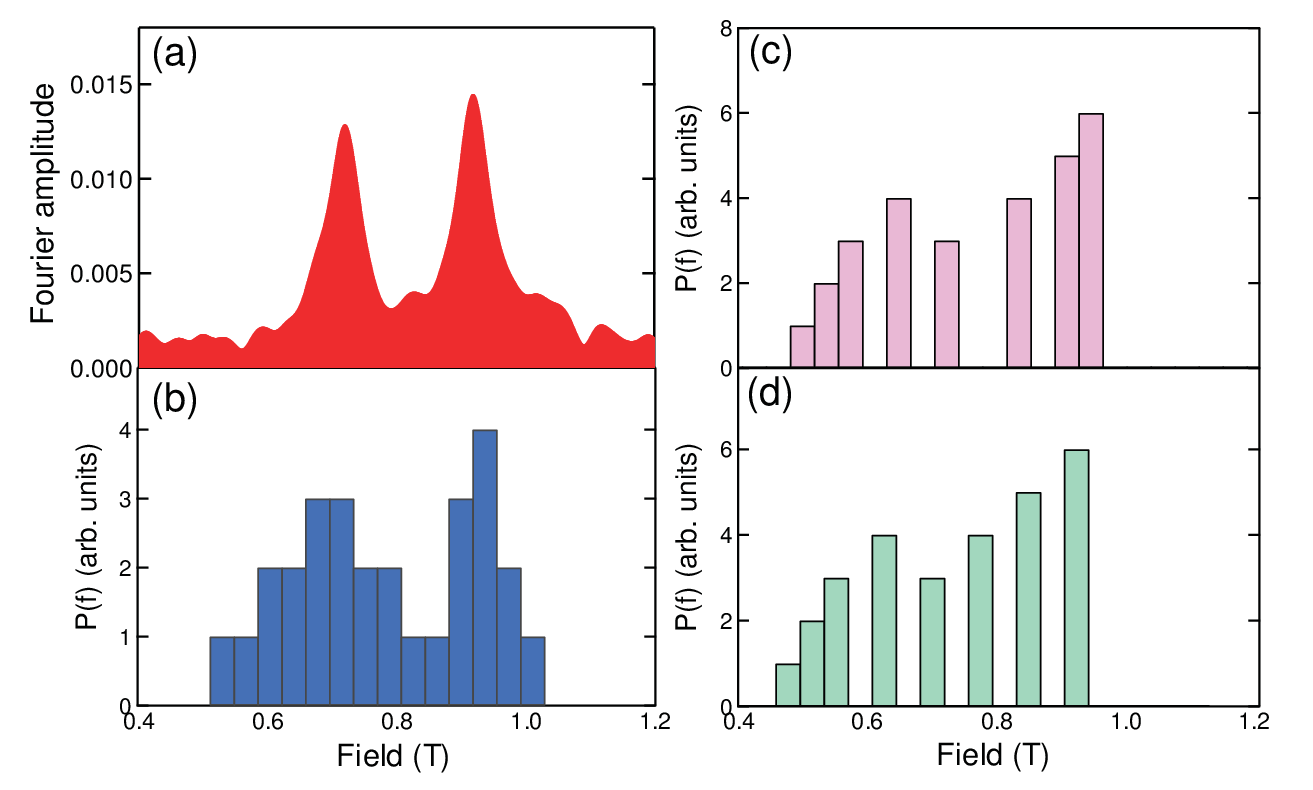}
	\caption{(a) Local field distribution ($B_{loc})$ at 2.3 K under a zero external field. (b) Simulated distribution of  $B_{\mathrm{loc}}$ at the 4$c$ site. Simulated distribution of  $B_{\mathrm{loc}}$ at the structurally-optimized site with a polaronic electron (not obtained from the DFT calculation) distributed on the nearest Mn ion only (c) or the nearby three Mn ions (d).}
	\label{Fig7}
\end{figure}
\par
\par
\begin{figure}[!t]
	\centering
	\includegraphics[width=0.75\linewidth,clip]{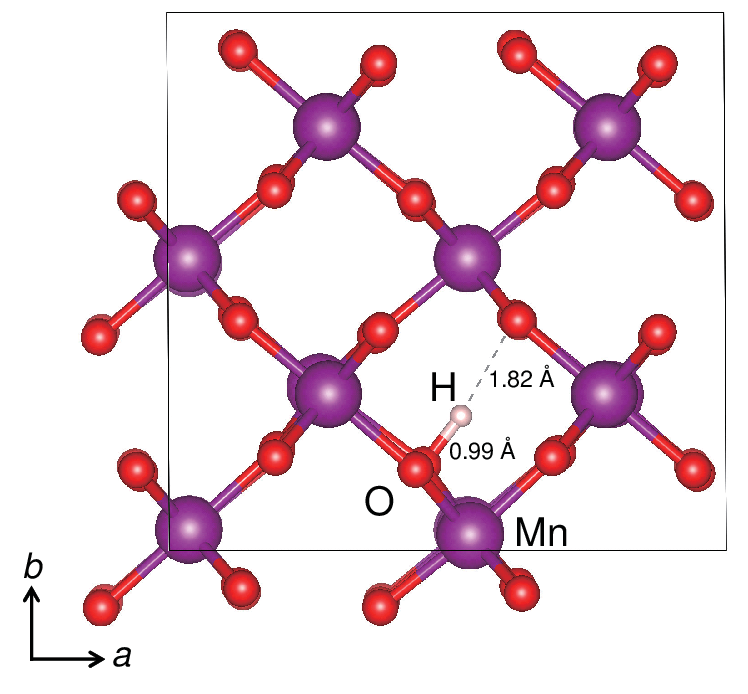}
	\caption{Deformation of the host lattice on hydrogen intercalation inferred from the DFT (VASP) calculation. The broken line represents the H\textperiodcentered \textperiodcentered \textperiodcentered O hydrogen bond.}
	\label{Fig8}
\end{figure}
\par
\subsection{Electronic and magnetic states}
Finally, we discuss the electronic and magnetic states of $\beta $-MnO$_{2}$ based on the results obtained so far. The formal magnetic moment of Mn$^{4+}$ ions should be 3$\mu_{B}$ as the localized spin system, because $\beta $-MnO$_{2}$ shows an insulating behavior in the low-temperature region ($T$ \textless 50 K). However, the Mn ordered moment is markedly reduced to 2.34$\mu_{B}$~\cite{Regls}, and this fact has been supported by the present $\mu $SR result as described above.
\par
The discrepancy in the size of the Mn moment may be attributed to the localized $e_{g}$ electrons originating from unintended impurity defects. If the Hund coupling is fairly large, $e_{g}$ electrons can become localized at a part of the Mn sites and perfectly polarized along the localized $t_{2g}$ moments at the expense of their itinerancy ~\cite{Sato}. However, massive amounts of impurities or oxygen defects would be required to induce such significant Mn moment reduction (3$\mu_{B} \to $ 2.34$\mu_{B})$. Moreover, Hund coupling usually enhances the magnetic moment, e.g., double-exchange interaction, thus it would lead to a contradictory result. Although there is a possibility that the delocalization effect of $t_{2g}$ electrons remains at low temperatures ~\cite{Sato}, we can rule it out because our $\mu $SR results clearly indicate that the local field distribution of $\beta $-MnO$_{2}$ is well reproduced by the summation of the dipolar interaction from localized Mn moments.
\par
Here, we propose an alternative scenario to explain the Mn moment reduction in $\beta $-MnO$_{2}$. Charge redistribution from a formal valence between a metal and its ligand is a fundamental process in transition metal oxides ~\cite{Boc}. For example, there are slightly-trivalent Cu ions (Cu$^{2+} \underline{L}$, $\underline{L}$: a hole on the oxygen ligand sites) in hole-doped high-$T_{C}$ superconductors ~\cite{LH} or oxygen holes in Co$^{3+}$/Co$^{4+}$ mixed valence state in Li$_{x}$CoO$_{2}$ ~\cite{Miz}. The degree of the charge transfer from O2$p$ orbital to transition-metal 3$d$ orbital depends on the energy difference between both orbitals, and also relates to their electronegativity. In the case of Mn$^{3+}$/Mn$^{4+}$ mixed-valent oxides, significant charge transfer between Mn and O ions has been observed in the x-ray Compton scattering ~\cite{Cmpt}. It is also in good agreement with the theoretical prediction that Mn--O bond is not purely ionic but partially covalent ~\cite{DV}. Such an intermediate character between ionic and covalent Mn-O bond may play an important role in the catalytic activity and the performance for electrochemical storage of manganese oxides.
\par
Based on these facts, we assumed the charge-transfer process from the O2$p$ orbital to the Mn 3$d$-$t_{2g}$ orbital in $\beta $-MnO$_{2}$. If the covalent character of Mn-O bond suppresses the increase of Mn$^{3+}$ ionic radius and stabilizes a low-spin electronic configuration ($t_{2g}^{4})$ absolutely, the Mn$^{3+}$ moment would be significantly reduced (2$\mu_{B})$ compared to that (4$\mu_{B})$ of a high-spin configuration ($t_{2g}^{3}e_{g}^{1})$. As a reference, a low spin state of Mn$^{3+}$ ion is often found in some tunnel-structure manganese oxides ~\cite{Shannon,LS}. The experimental value of the reduced Mn moment (2.34$\mu_{B})$ in $\beta $-MnO$_{2}$ can be reproduced by the charge transfer of about 0.1 electron from each of the surrounding six oxygen ions (or even less than this, if the oxygen spin polarization is antiparallel to Mn spins ~\cite{miya,ishii}). Further studies of each wave-function character, e.g., Compton scattering, will be required to support our argument. 

\section{Summary}
We have studied the electronic and magnetic state of $\beta $-MnO$_{2}$ in terms of hydrogen impurities by $\mu$SR technique combined with the first-principles calculations. We have found the following experimental results: (1) 85 {\%} of muons localize in the oxygen channels of the rutile structure as interstitial protons, (2) residual 15{\%} of muons are associated with lattice imperfection as Mn vacancies, and (3) 7.6{\%} of those muons (in the oxygen channels) form charge-neutral Mu state at 2.3 K, which suggests that hydrogen interstitials act as shallow donors whose ionization energy is less than 1 meV. Detailed analyses combined with the first-principles calculations showed that implanted muons may prefer the center of the oxygen channel to form a symmetric O-$\mu $-O bond due to a large isotope effect.

\section*{Acknowledgment}
This work was supported by the MEXT Elements Strategy Initiative to Form Core Research Centers, from the Ministry of Education, Culture, Sports, Science, and Technology of Japan (MEXT) under Grant No. JPMXP0112101001. The $\mu $SR experiments were performed under user programs (Proposal No. 2017A0113 and 2019MS02 at the Materials and Life Science Experimental Facility of the J-PARC, and M1712 at TRIUMF). We would like to thank the staff of J-PARC and TRIUMF for their technical support during the $\mu $SR experiment.


\end{document}